\documentclass[prl,aps,twocolumn,amsfonts,superscriptaddress,showpacs,floatfix]{revtex4}
\usepackage{color}
\usepackage{graphicx} 
\usepackage[tight]{subfigure}
\usepackage{multirow}
\usepackage{amsmath}
\newcommand{\ket}[1]{\ensuremath{\left|{#1}\right\rangle}}
\newcommand{\bra}[1]{\ensuremath{\left\langle{#1}\right|}}
\newcommand{\NS}{\ensuremath{\ket{\mathrm{NS}}}}
\newcommand{\SN}{\ensuremath{\bra{\mathrm{NS}}}}
\newcommand{\RK}{\ensuremath{\ket{{k}}}}
\newcommand{\RmK}{\ensuremath{\ket{{-k}}}}
\newcommand{\aD}{\ensuremath{\lvert\Delta\rvert}}
\begin{document}
\title{Understanding the entanglement entropy and spectra of 2D quantum systems through arrays of coupled 1D chains}
\author{Andrew J. A. James}
\author{Robert M. Konik}
\affiliation{CMPMS Dept., Brookhaven National Laboratory, Upton, NY 11973-5000, USA}
\date{\today}
\begin{abstract}
We describe an algorithm for studying the entanglement entropy and spectrum of 2D systems, as a coupled array of $N$ one dimensional chains in their continuum limit. Using the algorithm to study the quantum Ising model in 2D, (both in its disordered phase and near criticality) we confirm the existence of an area law for the entanglement entropy and show that near criticality there is an additive piece scaling as $c_{eff}\log (N)/6$ with $c_{eff} \approx 1$. \textcolor{black}{Studying the entanglement spectrum, we show that entanglement gap scaling can be used to detect the critical point of the 2D model. 
When short range (area law) entanglement dominates we find (numerically and perturbatively) that this spectrum reflects the energy spectrum of a single quantum Ising chain.}
\end{abstract}
\pacs{05.30.-d, 03.67.Mn, 64.70.Tg}
\maketitle

In the past decade concepts borrowed from information theory have become important tools in analyzing the properties of many-body quantum systems \cite{amico}.  The preeminent quantity in this regard is the bipartite entanglement entropy, $S_E$.  This measure of non-local quantum entanglement can be used to characterize quantum critical points \cite{vidal,holzney,calabrese_cardy}, access hidden topological order \cite{kitaev_preskill,levin_wen,fradkin_moore,balents}, and provides a simple measure for the applicability of the density matrix renormalization group (DMRG), one of the most commonplace numerical techniques in low dimensions \cite{vidal,schollwock,hallberg}.

Most is known about $S_E$ in one spatial dimension (1D).
In 1D $S_E$ signals the onset of criticality through an associated universal logarithmic divergence with system size, $L$ \cite{holzney,calabrese_cardy}.  
Both the coefficient of this divergence (i.e. the central charge of the theory's conformal algebra) and its subleading corrections in $L$ (determining the theory's operator content \cite{pasquier,calabrese}) serve to uniquely specify the underlying critical theory.

There is less certainty above 1D.
Regardless of criticality, $S_E$, possesses a term scaling as the area of the boundary separating the bipartite region \cite{eisert_review}.  Beyond this `area law' term, there can be subleading, universal contributions to $S_E$.
Generalizing the results in 1D, the AdS/CFT (Anti-de Sitter/Conformal Field Theory) correspondence suggests that $S_E$ in all
odd spatial dimensions will be characterized by universal logs \cite{ryu_takayanagi,nrt}.
Universal terms have also been argued for in the set of theories in 2D known as conformal quantum critical points (CQCP) 
\cite{fradkin_moore,hsu_fradkin,stephan1,oshikawa,zaletel,stephan2,fendley} as well as systems with spontaneously broken symmetries \cite{doyon,alba,metlitski_grover,metlitski_sachdev}. Recent studies of gapless states on the torus \cite{tag,ju,fendley} have confirmed the existence of apparently universal terms that depend on system shape.
\begin{figure}
\begin{center}
\includegraphics[width=0.28\textwidth]{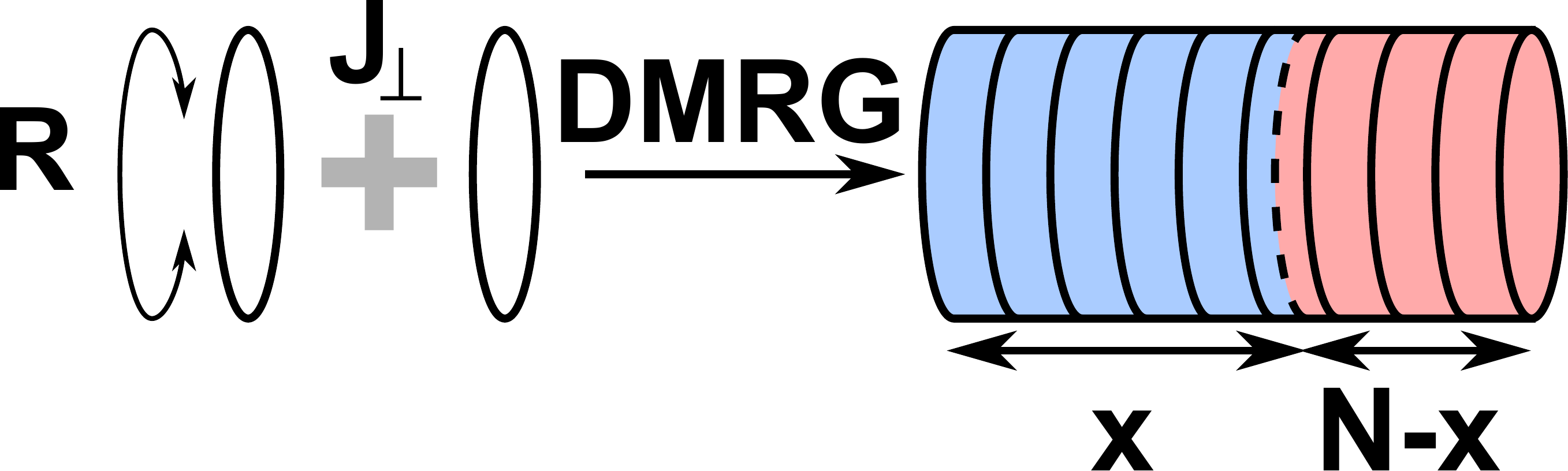}  {\large a)}\\
\includegraphics[width=0.28\textwidth]{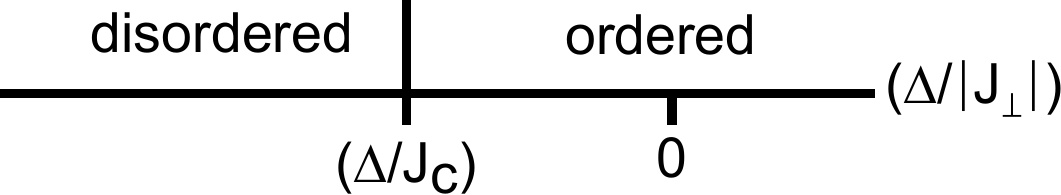}  {\large b)}
\end{center}
\caption{a) An array of continuum chains of length, R, with interchain
coupling, $J_\perp$. Each chain is a single site in our DMRG algorithm.
b) The phase diagram of coupled quantum Ising chains (QICs).}
\label{fig:setup}
\end{figure}

In this letter we demonstrate an algorithm for investigating the behavior of $S_E$ in 2D systems. The algorithm works by treating the model in an anisotropic limit: as a mixture of continuum and discrete degrees of freedom, making it amenable to a 1D-like DMRG algorithm. A major strength of the DMRG approach is that it works directly with the eigenvalues of the reduced density matrix $\rho_r$, the distribution of which governs the entanglement content. This allows us to simultaneously probe a variety of entanglement measures in 2D. Alternative techniques, such as Quantum Monte Carlo (QMC), that do not have direct access to $\rho_r$, are more restricted.

As a case study, we consider the quantum Ising model (or transverse field Ising model) in two spatial dimensions: a paradigmatic model for strongly correlated physics. We find that $S_E$ in 2D shares a number of characteristics with 1D -- including logarithmic scaling at criticality, in agreement with other studies \cite{tag,ju,fendley} -- and propose a scaling form to explain our results.

Furthermore, we analyse the entanglement spectrum (ES). While the ES was first studied as a means to understand the
efficacy of the DMRG algorithm \cite{okunishi,peschel1,peschel2} and then as a means of detecting topological order \cite{li_haldane}, it is
now being explored in non-topological systems to discover connections between the ES and the system's ordinary excitation spectra \cite{poilblanc,lauchli,lauchli2,haque,peschel3}.  We find that the scaling of the entanglement spectrum can be used to detect the critical point of the 2D system (a significant numerical advantage over calculating the energy gap), a result previously suggested only in 1D \cite{dechiara,lepori}.

\noindent {\bf Model and DMRG algorithm:}
We study the 2D quantum Ising model as a set of 1D quantum Ising chains (QICs), each with periodic boundary conditions and of length R, coupled through their spin operator:
\begin{equation}
H = \sum_i H_i^{1D~QI} + J_\perp\sum_{\langle ij\rangle}\int^{R}_0 dx\sigma_i(x)\sigma_j(x),
\end{equation}
where $i$ is a sum over chain index.  The Hamiltonian $H_i^{1D~QI}$ is taken in its continuum limit,
that of a massive Majorana fermion, 
$H_i^{1D~QI}= \int dx (i\bar\psi\partial_x\bar\psi -i\psi\partial_x\psi +i\Delta\bar\psi\psi)$
where $\bar\psi/\psi$ are right and left moving components of the Majorana fermion.
In lattice notation, $H^{1D~QI}=-J\sum_j\sigma^z_j\sigma^z_{j+1}+(g+1)\sigma^x_j$, and 
we identify $\sigma^z\rightarrow \sigma$, $\Delta = gJ$.
In this letter we focus on systems built from chains with negative mass ($\Delta<0$), as in this case one can drive the system to the critical point by increasing the magnitude of the interchain coupling, $J_\perp$ (see Fig. \ref{fig:setup}).
We employ a DMRG algorithm adapted to studying coupled 1D chains as described in \cite{ka2}, treating individual chains as equivalent to individual lattice sites in a conventional DMRG algorithm. As with all DMRG algorithms, this allows us to readily extract the entanglement entropy and spectrum.

This methodology is based in part on the truncated spectrum approach (TSA) to studying perturbed conformal and integrable field theories \cite{zamo}.
In the TSA the underlying conformal or integrable theory provides a particularly apt basis in which to study relevant (in the renormalization group sense) perturbations.  With such perturbations, the low energy sector of the full theory can be understood as a mixing of the low energy sector of the unperturbed theory (even if the two energy spectra are dramatically different).  Thus the high energy part of the theory can be neglected by imposing an energy cutoff or taken into account in a variational scheme borrowed from the numerical renormalization group (NRG) \cite{ka1}.  In the following we use a cutoff, $\Lambda$, on the chains.

As with perturbations in the TSA, our DMRG trades on the ability to compute matrix elements of the interchain coupling exactly, i.e. $\langle s|\sigma_j\sigma_{j+1}|s'\rangle$, where $|s\rangle,|s'\rangle$ are two states on a pair of neighbouring chains.  Using integrable or conformal 1D chains means we are able to incorporate much of the strongly correlated physics before the numerical analysis has even begun.

DMRG algorithms in 2D are limited relative to 1D because $S_E$ grows with the length of boundary between blocks (in this implementation, $R$, the chain length) \cite{schollwock,hallberg}. Approaching the thermodynamic limit then requires extrapolation and judicious choices of aspect ratio and boundary conditions \cite{stoudenmirewhite}.
Our use of continuum chains plays an important role in allowing the DMRG algorithm to work successfully.  In continuum
field theories, the finite size errors are exponentially suppressed in system size, R \cite{KM}.  This means that the chains
can be in the thermodynamic limit \textcolor{black}{for certain quantities,} even for comparatively small $R$ \cite{supp_mat}.  Keeping R small allows for smaller $S_E$ and hence an efficient DMRG implementation.  As a corollary to this we find we need to keep comparatively few eigenstates of the reduced density matrix, ranging from the 10's deep in the ordered phase to $\sim 200$ close to the critical point in order to obtain truncation
errors on the order of $10^{-5}$ (for additional details see \cite{supp_mat}).  Thus while the Hilbert space of 
the individual sites (i.e. the chains) can involve many hundreds of states, we need to keep far fewer states from the reduced density matrix.

The algorithm has been shown to successfully analyze various conventional properties of large arrays of coupled QICs \cite{ka2}.  It was able to reproduce the scaling form (in terms of the dimensionless combination $J_\perp^{4/7}\Delta^{-1}$) of the first excited gap in the disordered phase.  For this system it is possible to analytically compute the finite chain $R$ corrections, and an excellent match with the DMRG numerics was found.  Most significantly, the exponent $\nu$ governing the vanishing of the mass gap, $\Delta_{2D}$, as the critical coupling is approached, i.e. $\Delta_{2D}\sim|J_\perp-J_c|^\nu$, was computed. It was found to be $\nu=0.622\pm0.019$, in good agreement with the accepted value, $\nu=0.630$ and with a tensor based method \cite{xie,meurice}, for the 3D classical Ising model, demonstrating that the method successfully captures the physics of the 2D quantum system.

\begin{figure}
\includegraphics[width=0.4\textwidth]{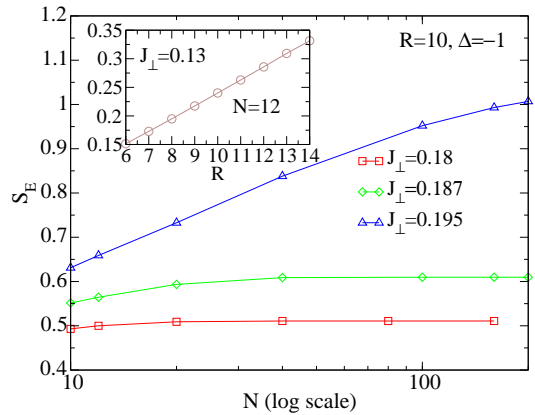}
\caption{$S_E$ as a function of the number of chains, N, in the disordered phase and near criticality. 
Inset: $S_E$ for the same phase as a function of R, the chain length.}
\label{fig:area_law}
\end{figure}
\noindent {\bf Entanglement Entropy:}
{ We couch our results for $S_E$  in terms of a scaling form applicable in the vicinity of criticality.  This scaling form must take into account the universal and the non-universal (i.e. cutoff, $\Lambda$, dependent) contribution to the area law as well as a subleading
universal logarithmic contribution that we believe we have detected in our numerics.  We allow this log to be dependent on the system aspect ratio (akin to that for 2D CQCP \cite{fendley})
- a necessity if we insist} 
{ on matching perturbation theory away from criticality. The scaling form we adopt is then
\begin{equation}
S_E = \alpha R\Lambda+\frac{c}{6}\log \frac{N}{\aD R} + \frac{R}{\xi_\parallel}f\Big( \frac{N}{\xi_\perp},\frac{ R}{\xi_\parallel},\frac{\xi_\perp}{\aD \xi_\parallel} \Big),
\end{equation}
where $\alpha$ is a non-universal constant, $ \xi_\parallel$ and $\xi_\perp$ are the correlation lengths parallel and perpendicular to the chains, respectively, and $f$ is a scaling function.
This form is chosen so that near criticality, where the scaling function is $f(0,0,\mathrm{const.})$, we obtain
\begin{align}
S_E \sim \alpha R \Lambda+ \frac{c}{6}\log \frac{N}{\aD R} + \mathrm{const.}
\label{eq:entropy_crit}
\end{align}
Here $\Delta$ serves as an effective inverse lattice spacing.
Far from criticality a perturbative calculation \cite{supp_mat} shows
\begin{equation}
S_E=\frac{1}{8}\frac{\aD R}{\xi_\perp}\exp \Big( -2\frac{\aD \xi_\parallel}{\xi_\perp}\Big),
\end{equation}
indicating that the non-universal contribution has vanished (i.e. $\alpha=0$) and that in this limit we can identify $f \sim -(c\xi_\parallel/6R) \log (N/(\aD R)) + g(\xi_\perp/(\aD \xi_\parallel))$ .}

{ In presenting our numerical results we start with
disordered chains, $\Delta=-1,\Lambda=8.0$ and display in Fig. \ref{fig:area_law} the behavior of $S_E$ as a function of $J_\perp,R$ and $N$, both deep in the disordered phase and close to criticality.} At very small $J_\perp$, $S_E$ displays an area law $S_E \sim R$ as expected, and is nearly constant with $N$ for values greater than $\sim 10$.
On increasing $J_\perp$ with fixed $R$, the value of $N$ at which $S_E$ saturates increases, and eventually a logarithmic dependence on $N$ becomes evident. At $J_\perp =0.195, R=10$ this $S_E \sim \log N$ behavior extends up to systems of $N \sim 100$.

\begin{figure}
\includegraphics[width=0.45\textwidth]{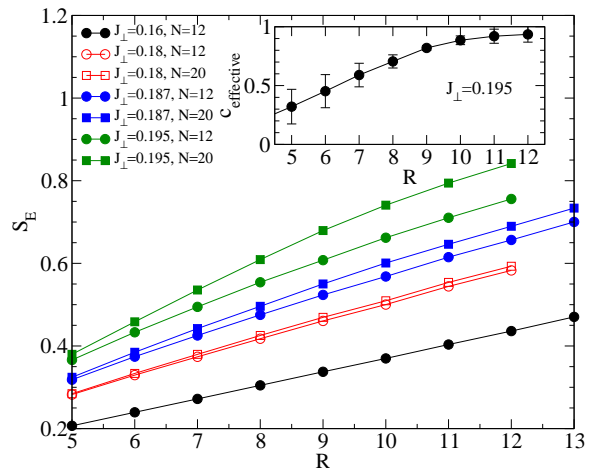}
\caption{$S_E$ closer to criticality for a block of length $N/2$ as a function of the chain circumference, $R$. For $J_\perp=0.187$ and $0.195$ the data sets are improved using truncation error extrapolation. Inset: The effective value of $c$ extracted. A power law fit for $R>7$ yields $c(R)=1.04(5)-[R/5.4(2)]^{-3.0(7)}$.}
\label{fig:area_law_crit}
\end{figure}

In Fig. \ref{fig:area_law_crit} we show the emerging $\log N$  behavior as a function of $R$. 
For $J_\perp=0.187,0.195$ the gradient becomes $N$ dependent, indicating a crossover as $\xi_\perp$ grows. Also visible is a crossover with $R$ between limiting forms of $f$.
\textcolor{black}{At these values of $J_\perp$ we find considerable improvement in our results on performing truncation error extrapolation \cite{white2007}, to compensate for the effect of the maximum DMRG correlation length \cite{nishino1996,stoudenmirewhite}.}
For $J_\parallel=0.195$, where the $\log$ behavior extends over the largest range, we can subtract Eq. (\ref{eq:entropy_crit}) at varied $N$ and fixed $R$ to estimate $c$.
Using data for $N=4,12,$ and $20$ we find the behavior in the inset of Fig. \ref{fig:area_law_crit}: the extracted value of $c$ tends to a constant for large $R$. A power law fit yields $c \to 1.04(5)$.

\begin{figure}
\includegraphics[width=0.45\textwidth]{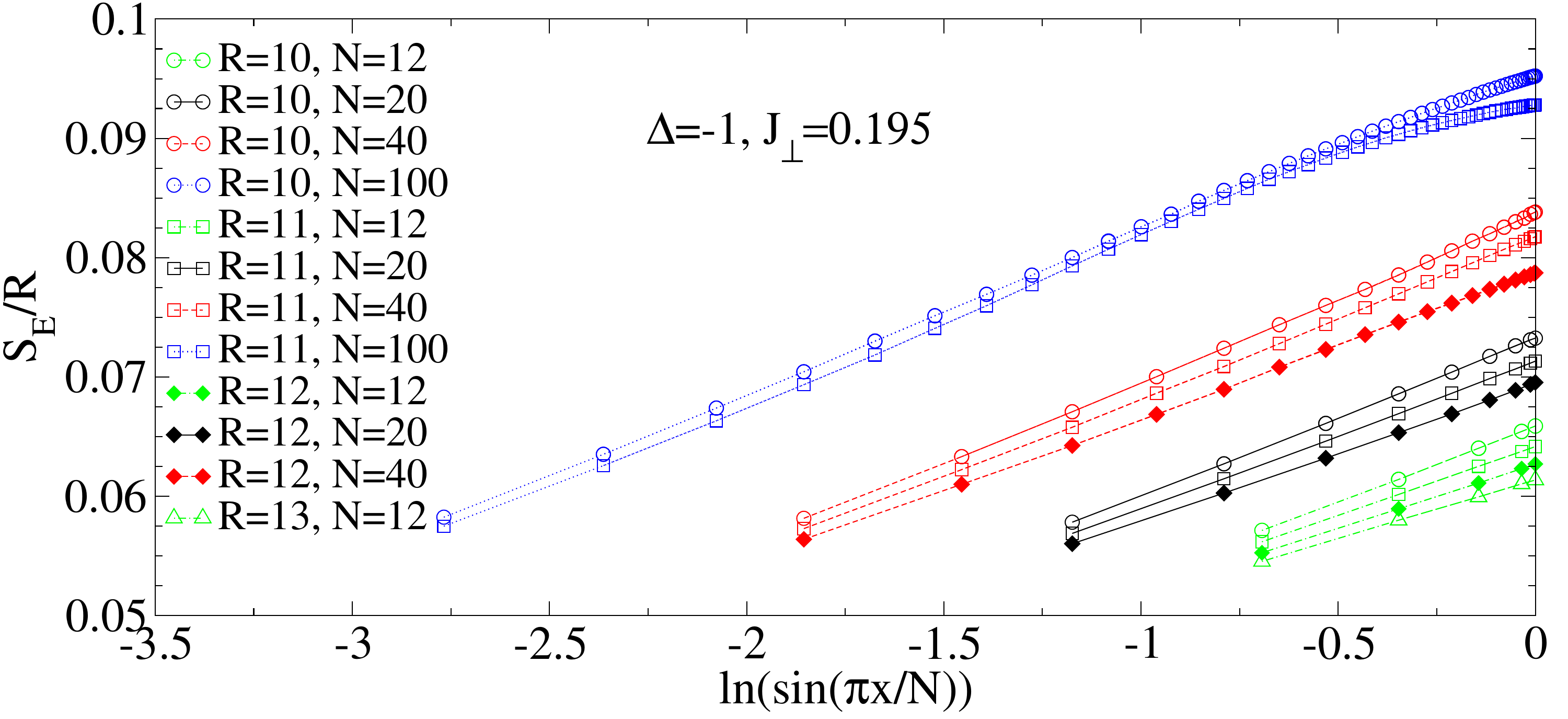}
\caption{$S_E/R$ for $J_\perp=0.195$ as a function of block length.} \label{fig:chord_scaling}
\end{figure}

Prompted by the logarithmic piece and in analogy with the 1D case, we look for `chord' scaling near criticality:
\begin{align}
S_E=\frac{c}{6} \log\Big[ \sin \Big( \frac{\pi x}{N}\Big)\Big]+\cdots
\end{align}
In our DMRG calculations $x$ corresponds to the number of chains in the `system' block while there are $N-x$ chains 
in the `environment' block (Fig. \ref{fig:setup}). In Fig. \ref{fig:chord_scaling} we plot $S_E$ 
versus the `chord length' at $J_\perp=0.195$ for a variety of aspect ratios; there is a clear 
linear relation. { Similar scaling behavior in 2D was observed in Ref. [\onlinecite{ju}] although recent analytical results for 2D CQCP's \cite{fendley} suggest that for 2D CQCP's this chord scaling is merely an excellent first approximation.}
\begin{table}[htdp]
\caption{Values of $c$ from data in Fig. \ref{fig:chord_scaling}.}
\begin{center}
\begin{tabular}{cc|cccc|}
& &\multicolumn{4}{c|} {R} \\
 & &10&11&12&13 \\ \hline
\multirow{4}{*}{N} &12    & 0.76 & 0.76 & 0.77 & 0.77 \\
 &20 & 0.78 & 0.80 & 0.83 & --- \\
& 40 & 0.80 & 0.84 & 0.87 & --- \\
& 100 & 0.80 & 0.86 & --- & ---
\end{tabular}
\end{center}
\label{default}
\label{tab:fits}
\end{table}%

\noindent {\bf Entanglement Spectra:}
In principle much more information is encoded in the full spectrum of the reduced density matrix, $\rho_{r}$, than in the number $S_E=-\mathrm{Tr} \rho_{r} \log \rho_{r} $ alone.
\begin{figure}[t]
\includegraphics[width=0.4\textwidth]{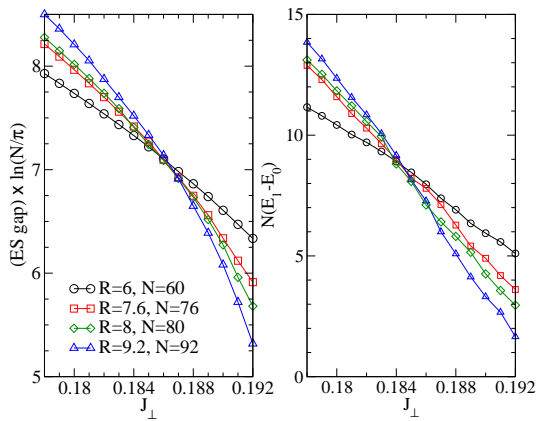}
\caption{Finite size scaling of both (left) the entanglement gap and (right) the real gap for DMRG calculations in which eigenvalues of $\rho_{r}$ are kept if they are $\ge 5\times 10^{-7}$.}
\label{fig:ent_scaling}
\end{figure}
The ES is usually defined in terms of a fictitious Hamiltonian, $\rho_r= \exp (-H_E)$ so that the ES `energies' are $\omega=-\log \rho_r$.
We first consider the entanglement `gap', $\Delta_{ES}$, the difference between the two lowest lying values of the ES.
\textcolor{black}{In Refs. \cite{dechiara,lepori} it was proposed that ES scaling in 1D can detect critical points. However these works do not find agreement with the scaling relation found by Calabrese and Lefevre \cite{calabrese_lefevre,pollmann_moore}:} for a 1D conformal system of finite length $N$, $\Delta_{ES}\sim\mathrm{const}/\log(N/\pi)$ \cite{calabrese_lefevre,pollmann_moore}. 

In Fig. \ref{fig:ent_scaling} we plot $\Delta_{ES} \log(N/\pi)$ against $J_\perp$ for a variety of 2D
systems with the same aspect ratio ($R/N$) but different sizes. Tellingly, the curves cross at a single point, $J_c=0.186(2)$ 
indicating that we can use this finite size scaling of the entanglement spectrum to discern the critical point of the 2D system. 
To support this claim we also perform finite size scaling on the true energy gap ($E_1-E_0$) in our system 
(right panel of Fig. \ref{fig:ent_scaling}) and find $J_c=0.185(2)$. This also agrees well with the RG improved value $J_c=0.184(3)$ in Ref. \onlinecite{ka2}. 
We note that calculating $E_1-E_0$ is considerably more difficult than $\Delta_{ES}$, as the former requires targeting the first 
excited state with the DMRG algorithm. Using $\Delta_{ES}$ to find $J_c$ therefore offers a significant numerical advantage.

\begin{figure}[t]
\includegraphics[width=0.45\textwidth]{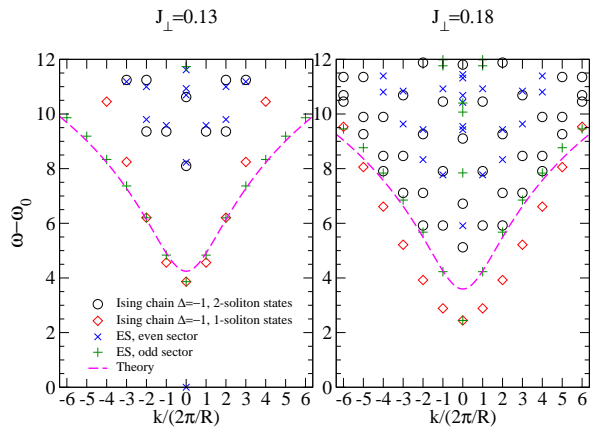}
\caption{Plot of the entanglement spectra both away from criticality (left), $J_\perp=0.13$ and closer
to criticality (right), $J_\perp=0.18$. $N=160$. In both cases the ES is measured relative to its lowest eigenvalue. The two sectors of the energy spectrum of a QIC are also plotted, rescaled so that the lowest band overlaps the lowest ES band at $k=0$. The curve labelled `theory' is a perturbative calculation in \cite{supp_mat}.}
\label{fig:ent_spectra}
\end{figure}
Finally we consider the ES as a function of the momentum, $k$, along the chain direction. It has been shown that the 
ES of spin ladders closely resembles the true energy spectrum of a single spin chain \cite{poilblanc,lauchli,lauchli2}. 
The spectrum of the QIC separates into two sectors, Neveu-Schwarz (NS) and Ramond (R) \cite{supp_mat}.
For a $\Delta<0$ chain these correspond to even and odd numbers of solitons along the chain respectively. 
Similarly the ES splits into two sectors, 
depending on whether the state has an even or odd number of chains in the NS sector
(assuming N/2 is even). 
Fig. \ref{fig:ent_spectra} shows that at $J_\perp=0.13$, far from criticality and where short range (area law) entanglement at the boundary 
dominates, the low lying ES resembles that of a single QIC, where the one and two soliton sectors are mimicked by
ES states with odd and even numbers of NS chains.
Closer to criticality, at $J_\perp=0.18$, the ES does not resemble that of a disordered QIC, in particular $\Delta_{ES} \to 0$.
A perturbative calculation for weak intrachain coupling gives $\omega =2\log (\Delta^2+k^2)+\mathrm{const}$ for the lowest `band' 
in the ES (see \cite{supp_mat}). The good agreement between this prediction (with $\Delta=-1$) and the $J_\perp=0.13$ 
spectrum is shown in Fig. \ref{fig:ent_spectra}.

The 1D like features that we see in our 2D system suggests the following interpretation.  Using the intuition that comes from
our anisotropic treatment of a 2D system, any 2D system can be thought of as a 
set of coupled continuum chains.  At a critical point, this anisotropic representation does not affect the critical properties (provided the critical point is a point and not a line where a lattice vs continuum treatment might control where along the line one ends up).
If at the 2D critical point, a finite number of chains become critical with the remaining chains massive with a gap of at least $\Delta_{min}$, one would expect to see 1D scaling.

In summary, we have shown that an unconventional DMRG technique can be used to study the entanglement content of strongly correlated 2D quantum systems. 
Using this technique we have established the existence of an additive logarithmic piece in $S_E$ with a universal coefficient $c \approx 1$ for the 2D quantum Ising model. We have also shown that the ES gap can be used to efficiently find a critical point in 2D and that when this gap is large and short range entanglement is dominant, the ES reflects the spectrum of a single Ising chain.

\noindent {\bf Acknowledgements}: This research was supported by the US DOE (DE-AC02-98CH10886).
We are grateful to P. Calabrese, F. Essler,  J.-M. St\'{e}phan and P. Fendley for helpful discussion and for the hospitality of the Galileo Galilei Institute at which part of this work was completed.
\section{Supplemental Material}
\noindent {\bf Details of Numerics:}
We consider arrays of $N$ chains with $N \le 200$ and $\Delta=-1$ (which sets the overall energy scale). 
The primary limitation on our numerics is storage for the states of two chains, necessary for the first basic step 
of the DMRG where we consider
a system of four chains. The number of these states grows rapidly with energy cutoff and chain length, $R$. 
In the following we set $E_c=8.0$ and $R\le12$ corresponding to the incorporation of $\lesssim 18000$ two chain states.

\begin{figure}[h]
\includegraphics[width=0.4\textwidth]{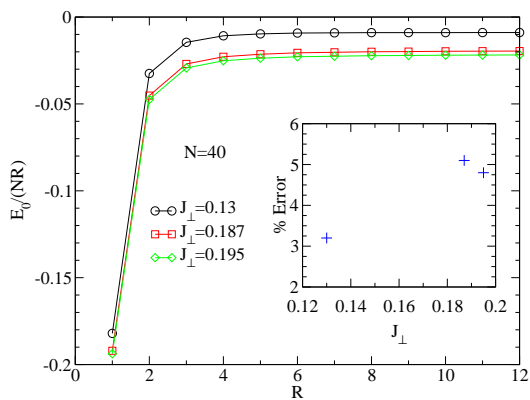}
\caption{The ground state energy per unit volume, as a function of $R$ for values of $J_\perp$ corresponding to deep in the disordered phase ($J=0.13$) and close to criticality ($J=0.187,0.195$). Inset: the percentage error at $R=6$ defined in terms of the ground state energy per unit volume at $R=12$. }
\label{fig:errors}
\end{figure}

\begin{figure}[h]
\includegraphics[width=0.4\textwidth]{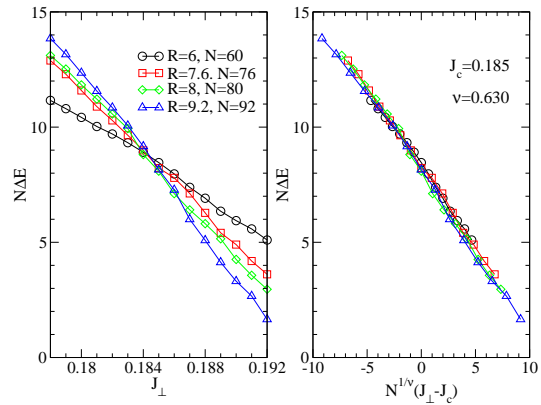}
\caption{Finite size scaling the real gap (left) and (right) data collapse assuming $\nu=0.60$ and $J_c=0.185$. Both for DMRG calculations in which eigenvalues of $\rho_{r}$ are kept if they are $\ge 5 \times 10^{-7}$.}
\label{fig:realgap_collapse}
\end{figure}

\begin{figure}[h]
\includegraphics[width=0.4\textwidth]{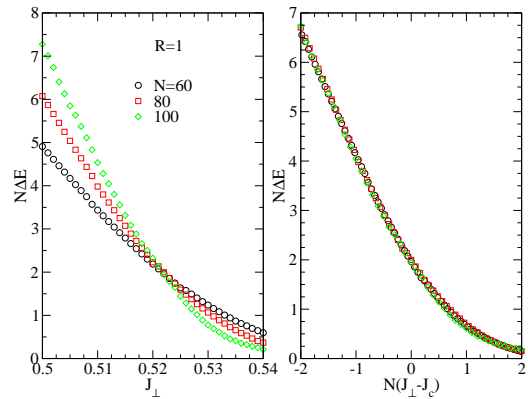}
\caption{Finite size scaling of (left) the real gap for $R=1$ and (right) collapse of the same data onto a single curve assuming $\nu=1$.}
\label{fig:R1_collapse}
\end{figure}

A number of tests of our method were carried out in Ref. \cite{ka2} including a check of the contribution from leading order finite size corrections. Here we discuss the role of chain length, $R$, for the parameters in this work.
In Fig. \ref{fig:errors} we plot the ground state energy, $E_0$ per unit volume as a function of $R$ for three different couplings, corresponding to deep in the disordered phase $J_\perp=0.13$ and much closer to criticality $J=0.187,0.195$. We see that there is rapid convergence as $R$ is increased, both in the disordered phase and near criticality, so that the change in $E_0/NR$ is $\lesssim 5\%$ between $R=6$ and $R=12$.
We also see a scaling collapse for $R\ge 6$ in Fig. \ref{fig:realgap_collapse} compatible with the 3D classical Ising exponent $\nu=0.630$. As described in Ref. \cite{ka2}, the leading cause of departure from this behaviour is the finite energy cutoff $E_c$, which slightly modifies $\nu$ (for $E_c=7.8$ it was found that $\nu=0.650$).
For small $R$ we expect to crossover to 1D quantum Ising critical behaviour. In Fig. \ref{fig:R1_collapse}, for $R=1$ and $E_c=20$, we show a scaling collapse with the 2D classical Ising exponent $\nu=1$.

\begin{figure}[h]
\includegraphics[width=0.4\textwidth]{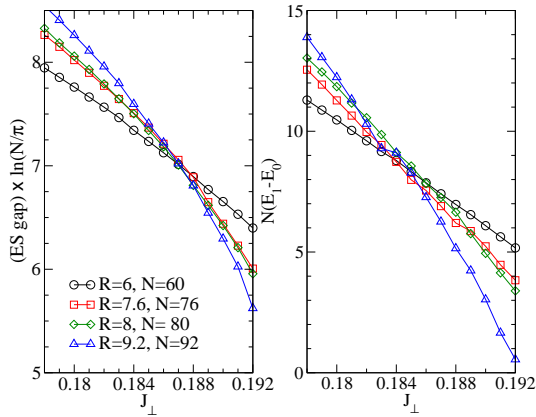}
\caption{Finite size scaling of (left) the real gap and (right) the entanglement gap for DMRG calculations in which eigenvalues of $\rho_{r}$ are kept if they are $\ge 25 \times 10^{-7}$.}
\label{fig:ent_scaling_larger}
\end{figure}

On approaching the critical point $S_E$ increases. This indicates that the spectrum of the reduced density 
matrix is no longer dominated by just a small number of eigenvalues.
Consequently the DMRG algorithm must retain a larger number of eigenvalues in order to accurately represent the target state 
with a larger computational penalty.
In our DMRG algorithm whether or not an eigenvalue is kept in each step is controlled by the condition $\rho_{red,i}>\rho_c$ 
where $\rho_c$ is a threshold value.
For the detailed plots of the energy gap, $E_1-E_0$, and $\Delta_{ES}$ close to criticality (Fig. 5 of the letter) 
we found it necessary to decrease $\rho_c$ by a factor of $5$ (effectively increasing the number of kept states by $50\%$) 
to produce sufficiently accurate results. In Fig. \ref{fig:ent_scaling_larger} we plot data obtained using the larger 
(less accurate) value of $\rho_c$. The effect is most clearly visible for the larger system sizes and especially for the 
plot of the energy gap, because the first excited state carries more entanglement relative to the ground state and is thus 
more challenging for the DMRG to represent, reducing the accuracy and rate of convergence. It was necessary to increase the 
number of convergence sweeps in the finite volume part of the DMRG routine from $\sim4$ for the ground state with $R=6, J_\perp=0.17$ 
to $\sim20$ for the first excited with $R=9.2, J_\perp=0.185$.
We also found it necessary to choose our system sizes such that $N/2$ was even, to avoid small even--odd effects associated with 
the number of chains in a block.
In Fig. \ref{fig:ent_scaling_larger} it is much harder to discern the crossing points, leading to crude estimates of $J_c$ as $0.187(3)$ 
and $0.185(5)$ from finite size scaling of $\Delta_{ES}$ and $E_1-E_0$ respectively.  We do see however from this computation
that as we increase the accuracy of the DMRG computation (as measured by kept states), the crossing points from the finite size
scaling of the ES and the gap to the first excited state move closer together.

\noindent {\bf Entanglement Spectra of Weakly Coupled Chains:}
Consider a system of $N$ disordered ($\Delta <0$) QICs, where $N/2$ is an even number, with weak intrachain exchange $J_\perp$.
The Hilbert space of a single QIC on a periodic interval of finite length $R$ (as summarized succinctly in \cite{fonseca})
splits into two sectors, 
known as Neveu-Schwarz (NS) and Ramond (R). For a disordered chain the states in the NS sector consist of even 
numbers of solitons with their momenta quantized as $k_i=2\pi m_i/R$ with $m_i$ taking half-integer values.
In contrast the states of the R sector consist of odd numbers of solitons with $k_i=2\pi n_i/R$ where $n_i$ takes integer values.
The energy of a $p$ soliton state is the sum $\sum_{i=1}^p \sqrt{\Delta^2+k_i^2}$ with $p$ even and odd for the NS and 
R sectors respectively (we neglect the two vacuum state energies, the difference between which is exponentially small in the length $R$).

The ground state of the $N$ chain system for $J_\perp=0$ is then a tensor product of zero soliton (vacuum) NS states.
\begin{align}
\ket{\psi_0}=\prod_{i=1}^{N} \NS_i.
\end{align}
Treating the intrachain coupling in the Hamiltonian as a perturbation
\begin{align}
H_\perp=J_\perp \int_0^R dr \sum_{i=1}^{N-1} \sigma_i(r) \sigma_{i+1}(r),
\end{align}
and using translational invariance along the chains, we find the first order contribution to the ground state is
\begin{align}
\ket{\psi_1}=&-J_\perp R \sum_{i=1}^{N-1}\Big(\prod_{j\ne i,i+1} \NS_j \Big) \nonumber \\
& \times \sum_k \frac{ \RK_i \RmK_{i+1} \bra{k} \sigma \NS \bra{-k}\sigma\NS  }{2\sqrt{\Delta^2+k^2}},
\end{align}
where we have used the property that matrix elements of the spin operator only connect different sectors. 
Furthermore we have assumed that the most important contribution comes from the one soliton (Ramond) states, $\RK$, 
as there is a mass gap, $\Delta$. We write this state as
\begin{align}
\ket{\psi_1}=&-J_\perp \sum_{i=1}^{N-1}\Big(\prod_{j\ne i,i+1} \NS_j \Big) \sum_k A_k \RK_i \RmK_{i+1}
\end{align}
where we have defined the function $A_k$.
For large $R$ the matrix elements simplify and
\begin{align}
A_k \approx \frac{\bar{\sigma}^2}{2\Delta^2}\Big(1+\Big[\frac{k}{\Delta}\Big]^2\Big)^{-1}
\end{align}
with $\bar{\sigma}=\lvert \Delta\rvert^{1/8} \times 1.35783834\cdots$ \cite{fonseca}. 
The second order contribution $\ket{\psi_2}$ is somewhat involved, but we will only need the piece that connects to the unperturbed ground state:
\begin{align}
\ket{\psi_{2,0}}=&-\frac{N-1}{2}J_\perp^2 \prod_{i=1}^{N-1} \NS_j \sum_k A_k^2.
\end{align}
The norm of the corrected state is
\begin{align}
\mathcal{N}&=\langle \psi_0 | \psi_0 \rangle +\langle \psi_1 | \psi_1 \rangle + \langle \psi_{2,0} | \psi_0 \rangle+\langle \psi_{0} | \psi_{2,0} \rangle+\mathcal{O}(J_\perp^3)\nonumber \\
& = 1 +\mathcal{O}(J_\perp^3).
\end{align}
We now form the density matrix
\begin{align}
\rho=\mathcal{N}^{-1}&(\ket{\psi_0}+\ket{\psi_1}+\ket{\psi_2})(\bra{\psi_0}+\bra{\psi_1}+\bra{\psi_2}) \nonumber \\
\end{align}
 and take the trace over the states of the $N/2$ chains on the right-hand side of the system, keeping terms to $\mathcal{O}(J_\perp^2)$ only.
The resulting contributions to the reduced density matrix, $\rho_r$, can be separated into even and odd sectors depending on whether they include an even or odd number of NS chains (equivalently an even or odd number of total solitons) on the leftmost $N/2$ chains.
Working in the even sector one finds a diagonal piece  
\begin{align}
\Big( 1-J_\perp^2\frac{N}{2}\sum_k A_k^2\Big)\prod_{i=1}^{N/2-1} \NS_i \SN_i, 
\end{align}
and off-diagonal terms
\begin{align}
&-J_\perp \sum_{i=1}^{N/2-1}\sum_k A_k \big(\ket{k}_i \ket{-k}_{i+1}\SN_i \SN_{i+1}+\mathrm{H.c.}\big)\nonumber \\
&\qquad \times \prod_{j\ne,i,i+1}^{N/2} \NS_j \SN_j +\mathcal{O}(J_\perp^2).
\end{align}
The resulting matrix can itself be diagonalised using second order perturbation theory (the off-diagonal $J_\perp^2$ terms would enter at $\mathcal{O}(J_\perp^4)$, hence their neglect above).
The lowest lying even soliton eigenvalue of $\rho_r$ is then given by
\begin{align}
\rho_r^0&= \Big( 1- \sum_k \big[J_\perp A_k\big]^2\Big)\nonumber \\
&=\Big( 1- \sum_k \Big[ \frac{J_\perp \bar{\sigma}^2}{2(\Delta^2+k^2)}\Big]^2\Big)\nonumber \\
&=\Big[ 1-\frac{\Delta R}{2}\Big(\frac{J_\perp \bar{\sigma}^2}{4\Delta^2} \Big)^2\frac{\Delta R +\sinh(\Delta R)}{\sinh^2(\Delta R/2)}\Big],
\end{align}
where we keep terms in $\rho_r^0 $ to $\mathcal{O}(J_\perp^2)$ only.
We extract the `ground state energy' of the ES as
\begin{align}\omega_0=&-\log \Big( 1-\frac{\Delta R}{2}\Big(\frac{J_\perp \bar{\sigma}^2}{4\Delta^2} \Big)^2\frac{\Delta R +\sinh(\Delta R)}{\sinh^2(\Delta R/2)}\Big) \nonumber \\
\approx &\frac{\Delta R}{2}\Big(\frac{J_\perp \bar{\sigma}^2}{4\Delta^2} \Big)^2\frac{\Delta R +\sinh(\Delta R)}{\sinh^2(\Delta R/2)},
\end{align}
which is of order $10^{-2}$ for the range of parameters we use.
In the odd sector the one soliton contributions can be organised by chain momentum, $k$:
\begin{align}
\rho_r^{1}(k)=\Big[ \frac{J_\perp \bar{\sigma}^2}{2(\Delta^2+k^2)}\Big]^2 
\end{align}
so that the lowest ES band is given by
\begin{align}
\omega_{1}(k)=-2\log \Big( \frac{J_\perp \bar{\sigma}^2}{2\Delta^2}\Big) +2 \log(1+(k/\Delta)^2).
\end{align}
This is the curve plotted in Fig. 6 of the letter.

We now consider corrections arising from the subleading $R$ dependence of $A_k$. 
Using the exact expression for the matrix elements \cite{fonseca} we find that the most important correction is at $2\pi n/R=k =0$:
\begin{align}
A_{0}=\frac{\bar{\sigma}^2}{2\Delta^2}\Big(1+\frac{e^{-\Delta R}(1-6\Delta R)}{(\Delta R)^{3/2}\sqrt{2\pi}} \Big)+\cdots
\end{align}
Corrections to the finite momenta contributions are much smaller, so we may continue to take
\begin{align}
A_{n \ne 0}=\frac{\bar{\sigma}^2}{2\Delta^2}\frac{1}{1+\Big[\frac{2\pi n}{\Delta R}\Big]^2}.
\end{align}
The two largest eigenvalues of the reduced density matrix are now
\begin{align}
\rho_r^0&= \Big( 1- \sum_k \big[J_\perp A_k\big]^2\Big)\nonumber \\
&=\Big( 1- \sum_{k\ne 0} \big[J_\perp A_k\big]^2\Big)\nonumber \\
&\quad +a^2\Big[1-\Big(1+\frac{e^{-\Delta R}(1-6\Delta R)}{(\Delta R)^{3/2}\sqrt{2\pi}} \Big)^2\Big]+\cdots\nonumber \\
&=1-a^2\Big[\frac{\Delta R(\Delta R +\sinh(\Delta R))}{8\sinh^2(\Delta R/2)}\nonumber \\
& \qquad \qquad +2\frac{e^{-\Delta R}(1-6\Delta R)}{(\Delta R)^{3/2}\sqrt{2\pi}}  \Big]+\cdots
\end{align}
and
\begin{align}
\rho_r^1(0)=a^2\Big(1+2\frac{e^{-\Delta R}(1-6\Delta R)}{(\Delta R)^{3/2}\sqrt{2\pi}} \Big)+\cdots
\end{align}
where we have defined $a=J_\perp \bar{\sigma}^2/(2\Delta^2)$.
We now use these expressions to calculate the entanglement entropy for large $R$ and small $J_\perp$.
The contribution to $S_E$ from $\rho_r^0$ is
\begin{align}
S_E(\rho_r^0)&=-\rho_r^0 \log \rho_r^0 \nonumber \\
&=a^2 \Delta R\Big[\frac{\Delta R +\sinh(\Delta R)}{8\sinh^2(\Delta R/2)}\nonumber \\
& \qquad \qquad +2\frac{e^{-\Delta R}(1-6\Delta R)}{(\Delta R)^{5/2}\sqrt{2\pi}}  \Big]+\cdots
\end{align}
To calculate the contribution from one soliton states we must approximate the sum over $n$ as an integral, taking into account the correction at $n=0$:
\begin{align}
S_E(\rho_r^1)&=-\sum_{n=-\infty}^{\infty} \rho_r^1(2\pi n/R) \log \rho_r^1(2\pi n/R) \nonumber \\
&=\sum_{n=-\infty}^{\infty}b_n = b_0+2\sum_{n=1}^{\infty}b_n
\end{align}
The $n=0$ term is
\begin{align}
b_0&=-2 a^2\Big[ \Big(1+2\frac{e^{-\Delta R}(1-6\Delta R)}{(\Delta R)^{3/2}\sqrt{2\pi}} \Big)\log a \nonumber \\
&\qquad  +\frac{e^{-\Delta R}(1-6\Delta R)}{(\Delta R)^{3/2}\sqrt{2\pi}}\Big]+\cdots.
\end{align}
The sum is
\begin{align}
2\sum_{n=1}^{\infty}b_n&=-a^2 \log a \Big [\frac{\Delta R(\Delta R +\sinh(\Delta R))}{4\sinh^2(\Delta R/2)}-2 \Big] \nonumber \\
&+4a^2\sum_{n=1}^{\infty}\Big(1+\Big[ \frac{2\pi n}{R}\Big]^2 \Big)^{-2}\log \Big(1+\Big[ \frac{2\pi n}{R}\Big]^2 \Big)+\cdots \nonumber \\
&=-a^2 \log a \Big [\frac{\Delta R(\Delta R +\sinh(\Delta R))}{4\sinh^2(\Delta R/2)}-2 \Big] \nonumber \\
&\qquad+\Delta R\frac{a^2}{2}(2\log 2-1)+\cdots
\end{align}
\begin{align}
S_E(\rho_r^1)=&- a^2\Big\{\log a \Big(4\frac{e^{-\Delta R}(1-6\Delta R)}{(\Delta R)^{3/2}\sqrt{2\pi}}  \nonumber \\
&\qquad+ \frac{\Delta R(\Delta R +\sinh(\Delta R))}{4\sinh^2(\Delta R/2)} \Big) \nonumber \\
&\qquad  +\frac{e^{-\Delta R}(1-6\Delta R)}{(\Delta R)^{3/2}\sqrt{2\pi}}\nonumber \\
&\qquad+\Delta R\Big(\frac{1}{2}-\log 2\Big)\Big\}+\cdots
\end{align}
The large $R$ limit for the lowest two entanglement spectrum bands gives an entanglement entropy,
\begin{align}
S_E=\frac{\Delta R}{4}\Big( \frac{J_\perp \bar{\sigma}^2}{2\Delta^2}\Big)^2 \Big[ -2\log \Big\lvert\frac{J_\perp \bar{\sigma}^2}{2\Delta^2}\Big\rvert-1 + 4\log 2 \Big].
\end{align}
We can also calculate the energy gap perturbatively.
To order $J_\perp$ this is
\begin{align}
\Delta_{2D}=\Delta \Big[ 1-4  \Big\lvert\frac{J_\perp\bar{\sigma}^2}{2\Delta^2} \Big\rvert \cos\Big( \frac{\pi}{N+1}\Big)\Big].
\end{align}

\end{document}